\documentclass[epj, nopacs]{svjour}

\usepackage{amssymb}  
\usepackage{amsmath}   
\usepackage{color}   

\newif\ifpdf
  \ifx\pdfoutput\undefined
  \pdffalse              % we are not running PDFLaTeX
\else
  \pdfoutput=1           % we are running PDFLaTeX
  \pdftrue
\fi

\ifpdf
  \usepackage[pdftex]{graphicx}
  \usepackage{thumbpdf}
  \definecolor{webgreen}{rgb}{0,.5,0} 
  \definecolor{webbrown}{rgb}{.6,0,0}
  \usepackage[pdftex,
  breaklinks=true,
  bookmarks=true,
  bookmarksopen=true,
  bookmarksnumbered=true,
  colorlinks=true, 
  linkcolor=webgreen, %defined below 
  filecolor=webbrown, %defined below 
  citecolor=webgreen,%defined below
  ]{hyperref}
  \hypersetup{
    pdfauthor   = {V.M.Ghete <Vasile.Ghete@cern.ch>},
    pdftitle    = {Measurement of Bs oscillations with the ATLAS detector},
    pdfsubject  = {ATLAS Scientific Note on Bs Oscillations},
    pdfkeywords = {ATLAS, B Physics, Scientific Note, Oscillations}
    }
  \DeclareGraphicsExtensions{.pdf}
\else
  \usepackage[dvips]{graphicx}
  \DeclareGraphicsExtensions{.eps,.ps,.eps.gz,.ps.gz}
  \newcommand{\texorpdfstring}[2]{#1}   
  \newcommand{\href}[2]{#2} 
\fi

\def\EPJC#1#2#3{{Euro.~Phys.~J.~C}{\bf #1}\ (#2)\ #3}
\def\PLB#1#2#3{{Phys.~Lett.~B}{\bf #1}\ (#2)\ #3} 
\def\PRD#1#2#3{{Phys.~Rev.~D}{\bf #1}\ (#2)\ #3} 
\def\NCimA#1#2#3{{Nuovo.~Cim.~A}{\bf #1}\ (#2)\ #3}
\def\CPC#1#2#3{{Comp.~Phys.~Comm.~}{\bf #1}\ (#2)\ #3} 
\def\NIMA#1#2#3{{Nucl.~Instr.~Meth.~A}{\bf #1}\ (#2)\ #3} 

\newcommand{\Bs}   {\ensuremath{B_{s}^{0}}}
\newcommand{\barBs}{\ensuremath{\bar{B}_{s}^{0}}}
\newcommand{\Bd}   {\ensuremath{B_{d}^{0}}}
\newcommand{\barBd}{\ensuremath{\bar{B}_{d}^{0}}}
\newcommand{\Ds}   {\ensuremath{D_{s}}}
\newcommand{\Dsm}  {\ensuremath{D_{s}^{-}}}

\newcommand{\Dm}   {\ensuremath{D^{-}}}

\newcommand{\aone} {\ensuremath{a_{1}}} 
 
\newcommand{\aonep}{\ensuremath{a_{1}^{+}}} 
\newcommand{\pim}  {\ensuremath{\pi^{-}}} 
\newcommand{\pip}  {\ensuremath{\pi^{+}}} 
\newcommand{\Km}   {\ensuremath{K^{-}}} 
\newcommand{\Kp}   {\ensuremath{K^{+}}}  
\newcommand{\Lb}   {\ensuremath{\Lambda_{b}^{0}}} 
\newcommand{\Lcp}  {\ensuremath{\Lambda_{c}^{+}}} 

\newcommand{\ifb}  {\ensuremath{{\mathrm{fb}}^{-1}}}

\newcommand{\ips}  {\ensuremath{{\mathrm{ps}}^{-1}}}

\newcommand{\dms}  {\ensuremath{\Delta m_s}}
\newcommand{\dmd}  {\ensuremath{\Delta m_d}}
\newcommand{\dgs}  {\ensuremath{\Delta \Gamma_s}}
\newcommand{\dgsg} {\ensuremath{\Delta \Gamma_s /\Gamma_s}}

\newcommand{\dmsSec} {\texorpdfstring{\boldmath $\Delta m_s$}{Delta-ms}}

\newcommand{\pT}   {\ensuremath{p_\mathrm{T}}}
\renewcommand{\deg}{\ensuremath{^\circ}}

\newcommand{\etal} {\hbox{et al.}}

\newcommand{\D}    {\displaystyle}

\newcommand{\1}    {\phantom{1}}

\begin{document}

\title{
Prospects for the measurement of {\boldmath \Bs}  oscillations 
with the ATLAS detector at LHC  
}              

\author{ B.~Epp, V.M.~Ghete, A.~Nairz}      

\institute{
Institute for Experimental Physics, 
University of Innsbruck,  Austria 
} 

\date{January 31, 2002} 

\abstract{
   The capabilities of the ATLAS detector to measure the  
$B_{s}^{0}$ oscillations in proton-proton interactions 
at the Large Hadron Collider  were evaluated. 
$B_{s}^{0}$ candidates in the $D_{s}^{-} \pi^{+}$
and $D_{s}^{-} a_{1}^{+}$ decay modes from semileptonic exclusive events 
were fully simulated and reconstructed, using a detailed detector
description. The sensitivity and the expected accuracy
for the measurement of the oscillation frequency were derived from 
unbinned maximum likelihood amplitude fits as functions of the
integrated luminosity. A detailed treatment of the systematic
uncertainties was performed. The dependence of the measurement 
sensitivity on various parameters was  also evaluated.

}

\maketitle

\section{Introduction}
\label{introduction}

The observed  \Bs\ and \barBs\ states are linear combinations of two mass 
eigenstates, denoted here as $H$ and $L$. Due to the non-conservation of 
flavour in charged weak-current interactions, transitions between 
\Bs\ and \barBs\ states occur with a frequency proportional to 
$\dms = m_H - m_L$.

Experimentally, these \mbox{\Bs -- \barBs} oscillations have not yet 
been observed directly.
In the Standard Model, their frequency is predicted in Ref. \cite{SM_dms} 
to be between 12.0 \ips\ and 17.6 \ips\ with 68\% CL, 
and lower than 20 \ips\ at 95\% CL, 
significantly larger than the corresponding value \dmd\ 
in the  \mbox{\Bd -- \barBd} system. 
From measurements done by the ALEPH, DELPHI and OPAL experiments at LEP, 
by SLD at SLC, and by CDF at the Tevatron, a combined lower bound of  
$\dms > 18.0$ \ips\
has been established at 95\% CL~\cite{dms_exp}. 
In the \mbox{\Bd -- \barBd} system, the oscillations have been directly 
observed and a rather precise value  of 
$\dmd = 0.472 \pm 0.017 $~\ips~\cite{PDG2000} has been measured.
 
The values for \dmd\ and \dms\ predicted in the Standard Model by computing 
the corresponding box diagrams, with the top-quark contribution assumed to 
be dominant, are proportional to $|V_{td}|^2$ and  $|V_{ts}|^2$ respectively. 
The direct determination of $V_{td}$ and $V_{ts}$ from  \dmd\ and \dms\ is, 
however, hampered by hadronic uncertainties. These uncertainties partially 
cancel in the ratio:
\begin{displaymath}
\frac{\dms}{\dmd} = \frac{M_{\Bs}}{M_{\Bd}} \;
                    \frac{\hat{B}_{\Bs} f_{\Bs}^2}{\hat{B}_{\Bd} f_{\Bd}^2} \;
                    \left | \frac {V_{ts}}{V_{td}} \right |^2 \; , 
\end{displaymath}
where $M_B$ are the $B$-meson masses, $\hat{B}_{B}$ are the bag parameters, 
and $f_B$ are the $B$-meson form factors. Using the experimentally-measured 
masses and a value for the ratio 
$\xi = \sqrt{\hat{B}_{\Bs}}f_{\Bs} / \sqrt{\hat{B}_{\Bd}}f_{\Bd}$  
which can be computed in lattice QCD, 
a better constraint for $V_{ts}/V_{td}$ can be obtained, which can then 
be converted into a constraint of $|V_{td}|$, the worst-measured side of 
the unitarity triangle.

In this note, an evaluation of the capability of the ATLAS detector to 
measure the \Bs\ oscillations in proton-proton interactions at the 
Large Hadron Collider (LHC) is presented. Some quantities involved in 
the measurement are still uncertain (cross sections, shape of the background, 
the final 
characteristics of the detector), therefore the dependence of the measurement 
sensitivity and of the expected accuracy on these parameters is also evaluated.
 
%\section{The ATLAS detector}
%\label{ATLAS}

\section{Event selection}
\label{EvSel}

The signal channels considered in this analysis for the measurement of 
\mbox{\Bs -- \barBs} oscillations are 
$\Bs \to \Ds \pi$  and $\Bs \to \Ds \aone$, 
with $\Ds \to \phi \pi$ followed by $\phi \to \Kp\Km$.

The event samples were generated using 
PYTHIA 5.7 \cite{PYTHIA}, passed then through the ATLAS full
GEANT-based  simulation 
program DICE (Inner Detector (ID) only) and reconstructed using an algorithm 
based on the Kalman filter implemented in the xKal\-man package from the ATLAS 
reconstruction program ATRECON. The physics model used for simulation, 
the description of the detector and of the reconstruction program 
are presented in detail in Ref.~\cite{PhysTDR}.

In the simulation, $b$-quark pairs were produced 
in $pp$-collisions at $\sqrt{s} = 14$~TeV by including direct production,
gluon splitting and flavour excitation processes for $b \bar{b}$ production. 
The $b$-quark was forced to decay semileptonically giving a muon with 
transverse momentum%
\footnote{The coordinate system has the $z$ direction along the beam
axis, with $x$-axis pointing to the centre of the accelerator ring and 
$y$-axis pointing upwards.  
The transverse momenta are computed with respect to the $z$ axis.} 
$\pT > 6$~GeV and pseudo-rapidity $|\eta| < 2.4$
which is used by the level-1 trigger to select the $B$ hadronic channels in 
ATLAS.
The associated $\bar{b}$ was forced to produce the 
required $B$-decay channels.  All the charged final-state particles
from the $B$ decay were required to have $\pT >0.5$~GeV and $|\eta| < 2.5$. 

The reconstruction of the \Bs\ vertex proceeded via the following steps 
(charge-conjugate states are implicitly included). 
The $\phi$ decay vertex was first reconstructed by considering all 
combinations of pairs of oppositely-charged tracks with $\pT > 1.5$~GeV 
for both tracks. Kinematic cuts on the angles between the two tracks  
$\Delta \varphi_{KK} < 10 \deg$  and  $\Delta \theta_{KK} < 10 \deg$ 
were also imposed. Here $\varphi$ denotes the azimuthal angle and
$\theta$ the polar angle in the coordinate system defined previously.
The two-track vertex was then fitted assigning the kaon 
mass to both tracks. Combinations passing a
fit-probability~\cite{FitProb} cut of 1\% 
with the invariant mass within $3\sigma_\phi$ of the nominal $\phi$ mass 
were selected as $\phi$ candidates. To all accepted $\phi$ candidates, a 
third negative track with $\pT > 1.5$~ GeV from the remaining ones was added.
The pion mass was assigned to the third track and a three-track vertex was 
refitted. Combinations of three tracks which had a fit probability greater 
than 1\% 
and an invariant mass within $3\sigma_{\Ds}$ of the nominal \Ds\  mass were 
selected as \Dsm\ candidates. 

For each reconstructed \Dsm\ meson, a search was made for \aonep\ candidates 
in three-particle combinations of the remaining charged tracks. 
In a first step, $\rho^0$ mesons were reconstructed from all combinations of 
two tracks with opposite charges and with $\pT > 0.5$~GeV for both tracks, 
each particle in the combination being assumed to be a pion. 
Kinematic cuts $\Delta \theta_{\pi\pi} < 15 \deg$ and 
$\Delta \varphi_{\pi\pi} < 35 \deg$  were used to reduce the combinatorial 
background. The two selected tracks were then fitted as originating from the 
same vertex; from the combinations passing a fit probability cut of 1\%, 
those with an invariant mass within 1.5~$\Gamma_{BW}$  of the nominal $\rho^0$
mass were selected as $\rho^0$ candidates.
Next, a positive track with $\pT > 0.5$~GeV from the remaining charged tracks 
was added to the $\rho^0$ candidate, assuming the pion hypothesis for the 
extra track. The three tracks were then fitted as originating from a common 
vertex, without any mass constraints. Combinations with a fit probability 
greater than 1\% 
and with an invariant mass within 300~MeV of the nominal \aone\  mass were 
selected as \aonep\ candidates. 
                
For the $\Bs \to \Dsm \pip$  channel, the \Bs\ decay vertex was reconstructed 
by considering all \Dsm\ candidates and adding a fourth track from the 
remaining tracks in the event. This track was required to have opposite charge
with respect to the pion track from the \Dsm\ and  $\pT > 1$~GeV. 
The four-track decay vertex was refitted including $\phi$ and  \Dsm\ mass 
constraints, and requiring that the total momentum of the \Bs\ vertex pointed 
to the primary vertex (within the primary vertex spatial resolutions of 
$\sigma_x = \sigma_y = 28$~$\mu$m and $\sigma_z = 46$~$\mu$m) 
and the momentum of \Dsm\ vertex pointed to the \Bs\ 
vertex.  
 
For the $\Bs \to \Dsm \aonep$ channel, the \Bs\ candidates were reconstructed 
combining the \Dsm\ candidates with the \aonep\ candidates. A six-track vertex
fit was then performed with mass constraints for the tracks from $\phi$ and 
\Ds; due to the large \aone\ natural width, the three tracks from the \aonep\
 were not constrained to \aone\ mass.  As in the $\Bs \to \Dsm \pip$ channel,
the total momentum of the \Bs\ vertex was required to point to the primary 
vertex and the momentum of \Dsm\  vertex was required to point to the \Bs\ 
vertex. 

In order to be selected as \Bs\ candidates, the four-track and six-track 
combinations were required to give a probability greater than 1\% 
for the vertex fit. The signed separation between the reconstructed \Bs\ 
vertex and the primary vertex, and between the \Dsm\ and \Bs\ vertex were 
required to be positive (the momentum should not point backward to the parent 
vertex). To improve the purity of the sample, further cuts were imposed: 
the accepted \Bs\ candidates were required to have a proper decay time greater
than 0.4 ps, an impact parameter smaller than 55~$\mu$m and $\pT > 10$~GeV.

Background to the channels being considered for the measurement of \dms\ can 
come from two sources: from other four- or six-body $B$-hadron decay channels, 
and from combinatorial background (random combinations with 
\linebreak %!!!!!!!!!!!!!!!!!!!!!!!!!!!!!!!!!!!!!!!!!!!!!!!!!!!!!!!!!!!!!!!!!!
some or all 
particles not originating from a $B$ decay).
For $\Bs \to \Dsm \pip$, the following four-body decay channels were 
considered as potential sources of background: $\barBd \to \Dsm \pip$, 
$\Bd \to \Dm \pip$  (with $\Dm,\; \Dsm \to \phi \pim$  and $\phi \to \Kp\Km$) 
and $\Lb \to \Lcp \pim$ followed by $\Lcp \to p \Km \pip$. 
The similar six-body decay channels considered as potential sources of 
background for $\Bs \to \Dsm \aonep$  were: $\barBd \to \Dsm \aonep$, 
$\Bd \to \Dm \aonep$ (with $\Dm,\; \Dsm \to \phi \pim$  and $\phi \to \Kp\Km$)
and $\Lb \to \Lcp \pim$ followed by $\Lcp \to p \Km \pip \pip\pim$.
The simulated four- and six-body background events were passed through the 
detailed detector-simulation program, reconstructed and analysed using the 
same programs, the same conditions and the same cuts as the signal events.

In order to study the combinatorial background, a very large sample of 
simulated inclusive-muon events is needed. The results presented here are 
based on a sample of 1.1 million $b\bar{b} \to \mu X$ events, with 
$\pT > 6$~GeV and $|\eta| < 2.4$ for the muon corresponding to the trigger 
conditions.

The $b\bar{b} \to \mu X$ sample was analysed in the framework of a 
fast-simulation program ATLFAST++~\cite{PhysTDR}, applying the same algorithms
and the same cuts that were used for the fully-simulated samples. 
A careful check was made of the performance of the fast-simulation program by 
running it on signal and  background samples, and comparing the 
results with those from the detailed simulation. Reasonable agreement was 
obtained 
for the number of reconstructed events and the widths of the mass peaks for 
the reconstructed particles.

A multi-level trigger is used to select the events for this analysis. The  
level-1 trigger is the inclusive muon trigger mentioned before. 
The level-2 trigger~\cite{LVL2_Ds} reconfirms the muon from level-1 trigger, 
then in an un-guided search for tracks in the Inner Detector reconstructs a 
$\phi$ meson and, adding a new track, a \Ds\ meson. The level-2 trigger uses 
dedicated online software.
63\%  of the signal events selected offline pass the level-2 trigger cuts; 
from the $b\bar{b} \to \mu X$ sample, $(3.4 \pm 0.2)\%$ of the events are 
selected. The level-3 trigger (the event filter) uses a set of loose 
offline cuts, reducing the rate to $(0.26 \div 0.41)\%$ of the $\mu X$ rate, 
depending on the actual values which are set for the cuts. 

\begin{figure*}[htbp]
  \begin{minipage}[t]{0.48\linewidth}
    \centering
    \includegraphics{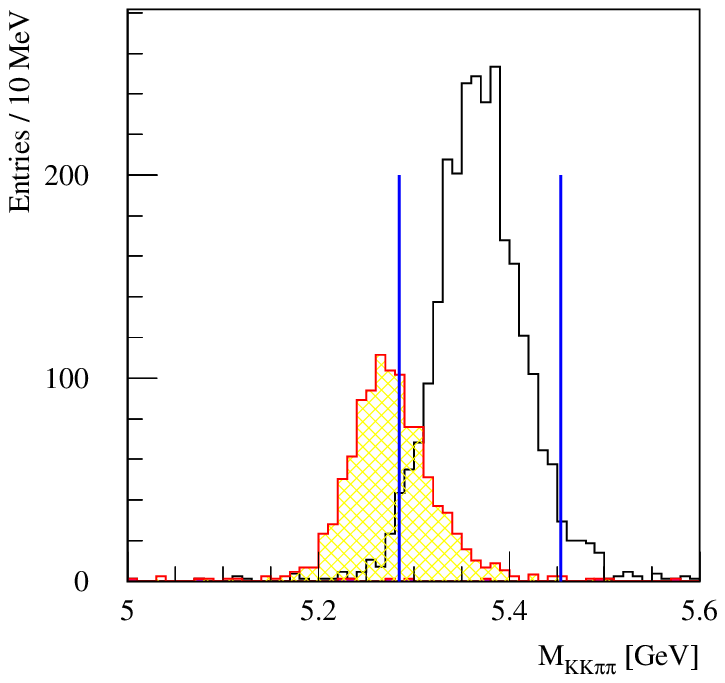}
    \caption{Reconstructed \Bs\ invariant-mass distribution for $\Bs \to \Dsm
      \pip$ decays. The open histogram shows the signal, the hatched
      histogram the background from $\barBd \to \Dsm \pip$ decays,
      and the dark histogram the fake reconstructed decays from
      the signal sample.  The combinatorial background is not shown
      here.}
    \label{fig:mass_Dspi}
  \end{minipage} \hfill
  \begin{minipage}[t]{0.48\linewidth}
    \centering
    \includegraphics{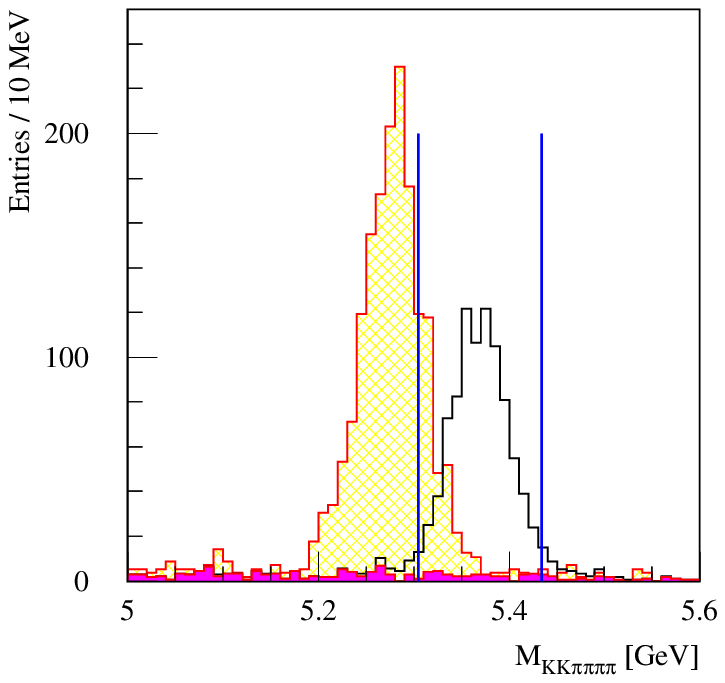}
    \caption{Reconstructed \Bs\ invariant-mass distribution for $\Bs \to \Dsm
      \aonep$ decays. The open histogram shows the signal, the hatched
      histogram the background from $\barBd \to \Dsm \aonep$ decays,
      and the dark histogram the fake reconstructed decays from
      the signal sample.  The combinatorial background is not shown
      here.}
    \label{fig:mass_Dsa1}
  \end{minipage}
\end{figure*}

\begin{table*}[htbp]
  \begin{center}
    \leavevmode
    \begin{tabular}{lccrrr}
      \hline
      \multicolumn{1}{c}{Process}       &
      \multicolumn{1}{c}{Cross-section} &
      \multicolumn{1}{c}{Effective } &
      \multicolumn{1}{c}{Simulated}     &
      \multicolumn{1}{c}{Rec.} &
      \multicolumn{1}{c}{Rec. events }
      \\  
                                   & 
      \multicolumn{1}{c}{(pb)} & 
      \multicolumn{1}{c}{cross-section (pb)} & 
      \multicolumn{1}{c}{events}   & 
      \multicolumn{1}{c}{events}   & 
      \multicolumn{1}{c}{for 10 \ifb}
      \\ \hline
$\Bs \to \Dsm \pip$      &   12.81   & 5.24 & 23893 & 2088 & 2370  \\ 
$\barBd \to \Dsm \pip$   &  \14.52   & 2.08 &  9989 &  369 &  400  \\ 
$\Bd \to \Dm  \pip$      &  \18.21   & 3.74 &  9989 &    2 &    3  \\  
$\Lb \to \Lcp (p \Km \pip) \pim$ 
                         &   89.44   & 2.27 &  9989 &    1 &    2  \\
$\Bs \to \Dsm \aonep$    &   12.81   & 2.08 & 18784 & 1512 &  870  \\ 
$\barBd \to \Dsm \aonep$ &   20.98   & 3.35 &  9699 &  190 &  340  \\ 
$\Bd \to \Dm  \aonep$    &  \18.20   & 1.40 &  9949 &    1 &    1  \\  
$\Lb \to \Lcp (p \Km \pip \pip\pim) \pim$ 
                         &  \11.97   & 0.20 & 10994 &    0 &    0  \\
Comb. background sum  &    & & $1.1 \times 10^6$ & see text    & 3750
      \\ \hline
    \end{tabular}
    \caption{Signal and background samples analysed for the
             study of \Bs - \barBs\  oscillations. The
             combinatorial  background is
             the sum for both $\Bs \to \Dsm \pip$ and 
             $\Bs \to \Dsm \aonep$ analysis channels.}
    \label{tab:nev}
  \end{center}
\end{table*}

The reconstructed \Bs\ invariant-mass distributions in the decay channels 
$\Bs \to \Dsm \pip$ and $\Bs \to \Dsm \aonep$ are shown in 
Figure~\ref{fig:mass_Dspi} and Figure~\ref{fig:mass_Dsa1}, respectively, 
for an integrated luminosity of 10~\ifb. 

The numbers of events expected for the various signal and background
channels that have been analysed are given in Table~\ref{tab:nev} for 
an integrated luminosity of 10~\ifb. 
The first cross-section column in Table~\ref{tab:nev} gives the
channel cross-section, without any cuts on final-state particles,
assuming a cross-section of 2.3~$\mu$b for the process
$b\bar{b} \to \mu (\pT > 6 \; \mathrm{GeV},\; |\eta| < 2.4)X$. 
The effective cross-section is the  cross-section after the cuts  
on charged final-state particles were applied during simulation.

The events reconstructed from the samples for the exclusive decay modes 
were counted in a $\pm 2 \sigma$  window around the nominal \Bs\ mass. 
Using the fraction of events reconstructed in the simulated sample and 
the number of events expected for an integrated luminosity of 10~\ifb, 
the expected number of reconstructed events was estimated. 
Corrections for muon efficiency (on average 0.82) and 
for level-2 trigger efficiency (0.63) were also applied.
 
A total of 3240 reconstructed events is expected for the $\Bs \to \Ds \pi$ 
and $\Bs \to \Ds \aone$  decay channels for an integrated luminosity of 
10~\ifb.

The only significant background comes from the $\barBd \to \Dsm \pip$ and 
$\barBd \to \Dsm \aonep$ channels, and from the combinatorial background. 
Note that the number of reconstructed events from the two  decay channels 
is conservative since the branching-ratio values used are upper limits. 
As expected, due to the combination of the \Dm\ mass shift 
($M_{\Dm}-M_{\Ds} = 90$~MeV) 
and  \Bd\ mass shift 
($M_{\Bs}-M_{\Bd} \approx 100$~MeV), 
very few $\Bd \to \Dm  \aonep, \; \Dm \pip$ events are reconstructed in a
$\pm 2 \sigma$  window around \Bs\  nominal mass. 
Due to the different decay topology, the $\Lb \to \Lcp \pim$ channel does 
not give any contribution to the background. 

The statistics available for estimating the combinatorial background are 
very limited, despite the large size (1.1 million events) of the $\mu X$ 
sample. Each simulated event was therefore passed 50 times through the 
fast-simulation program, different random smearing of the track parameters 
being applied each time. The number of background events was counted in 
a $2 \sigma$ mass window around the \Bs\ nominal mass. On average, 0.12 
events per pass were reconstructed in the mass window, summing the 
$\Bs \to \Dsm \pip$ and $\Bs \to \Dsm \aonep$ channels. 
Normalising to the number of $\mu X$ events expected for an integrated 
luminosity of 10~\ifb, applying correction factors for the reconstruction 
and trigger efficiencies, the combinatorial background was estimated to be 
3752 events, and the range of variation was estimated to be between 2079 and 
5425 events at 90\% CL. Correlations between the results from the 50 passes 
were taken into account.

\section{Proper-time reconstruction and resolution }
\label{pTime}

The proper time of the reconstructed \Bs\  candidates was computed from 
the reconstructed transverse decay length, $d_{xy}$, and from the \Bs\ 
transverse momentum, \pT:
\begin{displaymath}
t = \frac {d_{xy} M_{\Bs}} 
          {c \pT}
 \equiv  d_{xy} g
\end{displaymath}
where $g = M_{\Bs}/(c \pT)$.

The transverse decay length is the distance between the interaction point and 
the $b$-hadron decay vertex, projected onto the transverse plane. 
Figure~\ref{fig:bsdspi_dxy_rsd} shows, for the example of the  
$\Bs \to \Dsm (\phi \pim) \pip$ 
decay mode, the difference $d_{xy}-d_{xy}^0$ fitted with two Gaussian 
functions, where $d_{xy}^0$ is the true transverse decay length. 
For each event, the decay-length uncertainty, $\sigma_{d_{xy}}$, was estimated 
from the covariance matrices of the tracks associated with the vertices. 
The pull of the transverse decay length, $(d_{xy}-d_{xy}^0)/\sigma_{d_{xy}}$,
was found to have a Gaussian shape with a width of 
$S_{d_{xy}} = 0.959 \pm 0.017$ for the 
$\Bs \to \Dsm \pip$ channel and $S_{d_{xy}} = 0.954 \pm 0.020$ for the 
$\Bs \to \Dsm \aonep$ channel.

The distributions for $(g-g_0)/g_0$ also have a Gaussian shape for both 
\Bs\ decay channels, with a width of $S_g = (0.715 \pm 0.014) \%$ for the 
$\Bs \to \Dsm \pip$ channel and $S_g = (0.636 \pm 0.013) \%$ for the 
$\Bs \to \Dsm \aonep$ channel. Here $g_0=t_0/d_{xy}^0$, with $t_0$ being 
the true proper time. The $(g-g_0)/g_0$ distribution for the 
$\Bs \to \Dsm (\phi \pim) \pip$ decay mode is shown in 
Figure~\ref{fig:bsdspi_g_g0}. 

\begin{figure}[htbp]
  \begin{minipage}{7.5cm}
    \centering
    \includegraphics{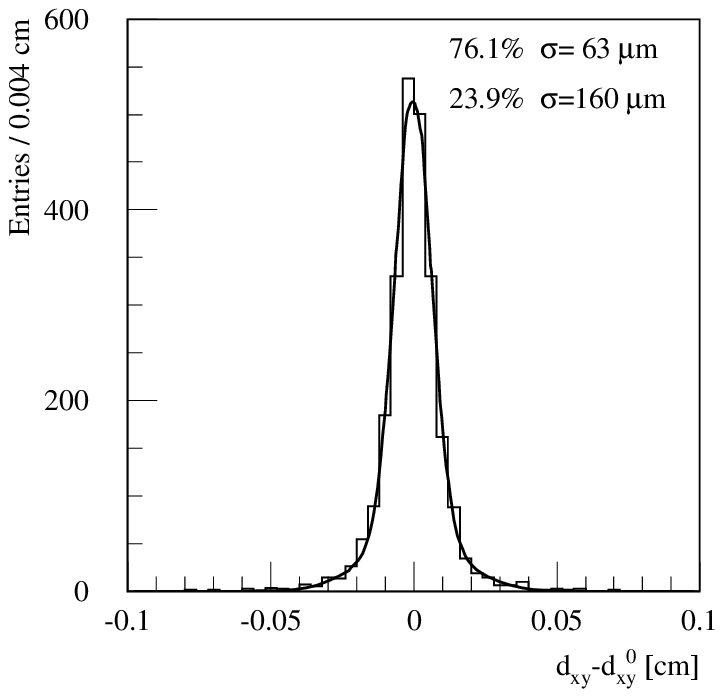}%
    \caption{Decay-radius resolution for the decay channel 
             $\Bs \to \Dsm (\phi \pim) \pip$.}
    \label{fig:bsdspi_dxy_rsd}
  \end{minipage} \hfill \vspace{2ex}
  \begin{minipage}{7.5cm}
    \centering
    \includegraphics{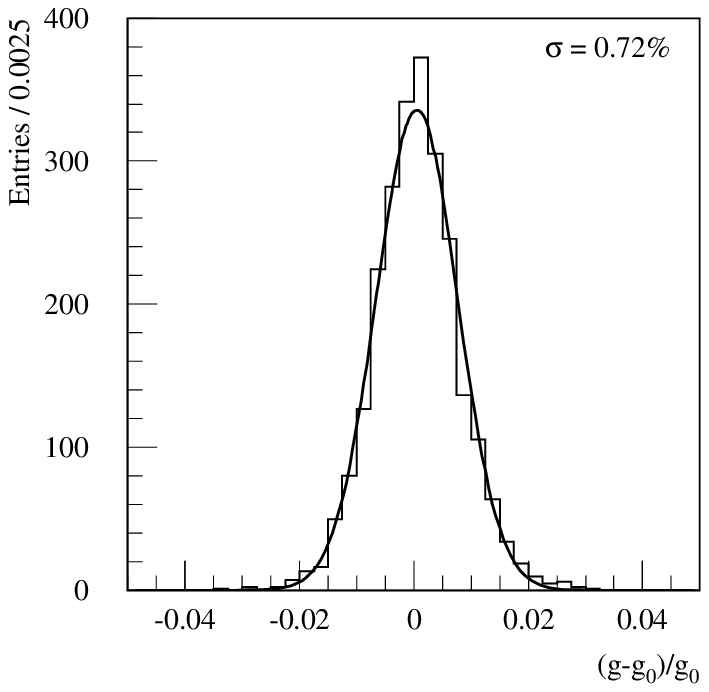}
    \caption{Fractional resolution on g-factor for the decay channel 
             $\Bs \to \Dsm (\phi \pim) \pip$.}
    \label{fig:bsdspi_g_g0}
  \end{minipage}
\end{figure}

The proper-time resolution function $\mathrm{Res}(t \, | \, t_0)$ was 
pa\-ram\-e\-trised 
with the sum of two Gaussian functions, with parameters given in 
Table~\ref{tab:ptres2G_par}:
\begin{equation*}
\begin{split}
    \mathrm{Res}(t \, | \, t_0)  = & 
                              f_{1} \frac{1}{\sigma_{1}\sqrt{2\pi}}\;\,
           \exp\left( -\frac{(t-t_0)^{2}}{2\,\sigma_{1}^{\; 2}}\right)
           + 
            \\ 
                                   & 
                              f_{2} \frac{1}{\sigma_{2}\sqrt{2\pi}}\;\,
           \exp\left( -\frac{(t-t_0)^{2}}{2\,\sigma_{2}^{\; 2}}\right)
\end{split}
\end{equation*}

\begin{table}[htbp]
\begin{center} 
  \begin{tabular}{ccccc} \hline
            &  \multicolumn{2}{c}{$\Bs \to \Dsm\,\pi^+$}   
            &  \multicolumn{2}{c}{$\Bs \to \Dsm\,a_1^+$}  \\ \cline{2-5} 
  $\alpha$  &  \multicolumn{1}{c}{Fraction $f_{\alpha}$}
            &  \multicolumn{1}{c}{Width $\sigma_{\alpha}$}
            &  \multicolumn{1}{c}{Fraction $f_{\alpha}$}
            &  \multicolumn{1}{c}{Width $\sigma_{\alpha}$}        \\ 
            &  (\%) & (fs) &  (\%) & (fs)               \\ \hline
1 & $59.6 \pm  6.6$  & $\phantom{1}51.5 \pm  4.0$ & 
    $62.5 \pm 14.1$  & $\phantom{1}51.6 \pm  \phantom{1}6.4$
                                                                  \\ 
2 & $40.4 \pm  6.6$  & $          107.3 \pm  8.5$ & 
    $37.5 \pm 14.1$  & $\phantom{1}92.8 \pm 12.7$ 
                                                                  \\ \hline
            &  \multicolumn{2}{c}{$\barBd \to \Dsm\,\pip$}   
            &  \multicolumn{2}{c}{$\barBd \to \Dsm\,a_1^+$}          \\ \hline
1 & $59.6 \pm  1.8$  & $\phantom{1}56.9 \pm  5.5$ & 
    $62.5 \pm  2.6$  & $\phantom{1}60.3 \pm  \phantom{1}7.2$ 
                                                                  \\ 
2 & $40.4 \pm  1.8$  & $          102.9 \pm  9.6$ & 
    $37.5 \pm  2.6$  & $          100.1 \pm 31.6$
                                                                  \\ \hline
  \end{tabular}
  \caption{Proper-time resolution function $\mathrm{Res}(t \, | \, t_0)$ 
           parametrization with the sum of two Gaussian functions.} 
  \label{tab:ptres2G_par}
\end{center}
\end{table}

Figures~\ref{fig:ptres_dspi} and ~\ref{fig:ptres_dsa1} show, for the decay 
channels $\Bs \to \Dsm \pip$ and $\Bs \to \Dsm \aonep$, the proper-time 
resolution together with the parametrization obtained from the  function given
above. The distributions show deviations from the Gaussian shape, illustrated 
by the significant fraction of the second, broader Gaussian function in the 
pa\-ram\-e\-tri\-za\-tion
function. For \Bd\ mesons, the ratio of the two Gaussian 
functions was fixed to the ratio from \Bs\ parametrization.
  
The parametrization chosen above reproduces well the tails seen in the 
distribution for reconstructed events. It has the advantage that the integrals
in the log-likelihood function given in Eq.~(\ref{fun:LL}) can be computed 
analytically, much
faster than the numerical computation. The dependence of the proper-time 
resolution on the proper time 
$\D \sigma(t_0)=\sqrt{(gS_{d_{xy}}\sigma_{d_{xy}})^2+(t_0S_g)^2}$
can be neglected in the first approximation, as the factor $S_g$ is very small.

\begin{figure}[htbp]
  \begin{minipage}[t]{7.5cm}
    \centering
    \includegraphics{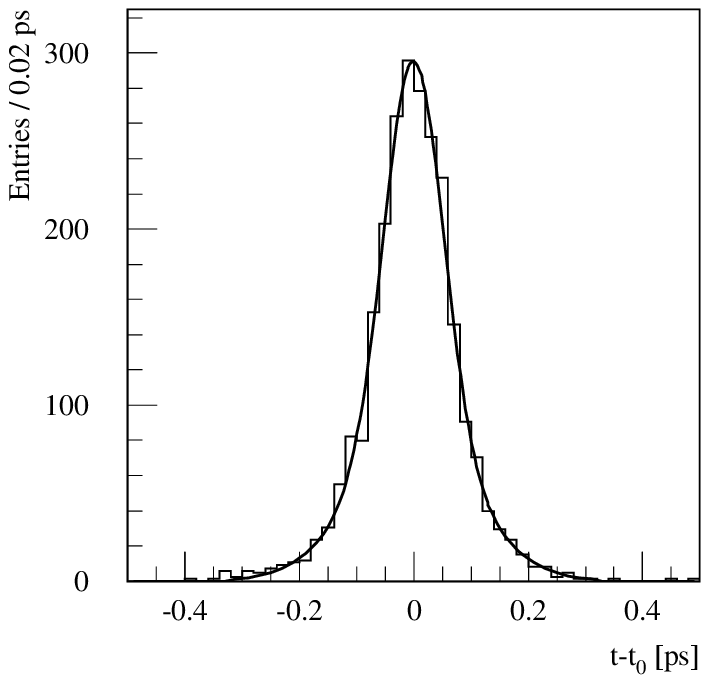} 
    \caption{Proper-time resolution for the decay channel $\Bs \to \Dsm \pip$.
             The curve displays the resolution as obtained from the  
             $\mathrm{Res}(t \, | \, t_0)$ function given in the text.}
    \label{fig:ptres_dspi}
  \end{minipage}  \hfill \vspace{2ex}
  \begin{minipage}[t]{7.5cm}
    \centering
   \includegraphics{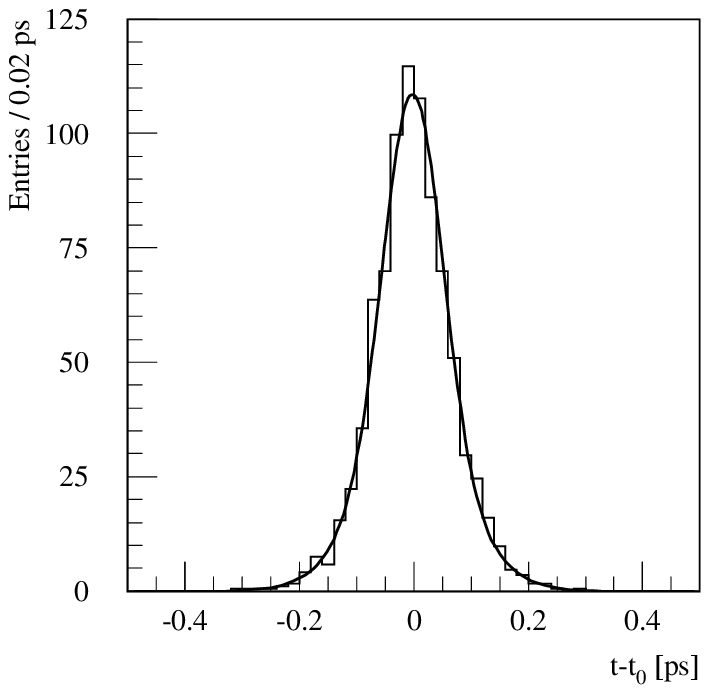}
    \caption{Proper-time resolution for the decay channel 
             $\Bs \to \Dsm \aonep$.
             The curve displays the resolution as obtained from the  
             $\mathrm{Res}(t \, | \, t_0)$ function given in the text.}
    \label{fig:ptres_dsa1}
  \end{minipage}
\end{figure}

\section{Likelihood function}
\label{LL}

The probability density to observe an initial $B_{j}^{0}$  meson ($j=d, \;s$)
decaying at time $t_0$ after its creation as a $\bar{B_{j}^{0}}$
meson is  given by:
\begin{equation}
  \begin{split}
    p_j(t_0, \; \mu_0)=  \D
    &
     \D \frac{\Gamma_{j}^{2} - \left( \frac{\Delta\Gamma_{j}}{2}\right)^{2}}
    {2\,\Gamma_{j}} \mathrm{e}^{- \Gamma_{j} t_0} \times \\
    &
    \left ( \cosh \frac{\Delta\Gamma_{j} t_0}{2} + \mu_0 \cos\Delta m_{j} t_0 
    \right) 
  \end{split}
  \label{eq:oscPr}
\end{equation}
where $\Delta\Gamma_{j}=\Gamma_\mathrm{H}^{j}-\Gamma_\mathrm{L}^{j}$, 
$\Gamma_j = (\Gamma_\mathrm{H}^{j} + \Gamma_\mathrm{L}^{j})/2$ and 
$\mu_0 = -1$. 
For the unmixed case (an initial  $B_{j}^{0}$ meson decays as a $B_{j}^{0}$ 
meson at time $t_0$), the probability density 
is given by the above expression with $\mu_0 = +1$. 
The small effects of CP violation are neglected in the above relation. 
Unlike $\Delta \Gamma_d$, which can be safely neglected, the width difference 
in the \mbox{\Bs -- \barBs} system  \dgs\ could be as much as 20\% 
of the total width~\cite{Bs_width_dif}.

The above probability is modified by experimental effects. The probability as 
a function of $\mu_0$ and the reconstructed proper time $t$ is obtained as
the convolution of $p_j(t_0, \; \mu_0)$ with the proper 
time resolution $\mathrm{Res}_j(t \, | \, t_0)$: 
\begin{equation*}
  q_j(t, \mu_0)= N 
  \int_{t_\mathrm{min}}^\infty  
       p_j(t_0,\mu_0)\; \mathrm{Res}_j(t\; |\; t_0) \; \mathrm{d}t_0 
\end{equation*}
with $N$ a normalization factor.
Assuming a fraction $\omega_j$ of wrong tags at production or decay, the 
probability becomes:
\begin{equation}
  q_j(t, \mu)= (1-\omega_j) q_j(t,\mu) + \omega_j q_j(t,-\mu) 
\label{eq:qjt}
\end{equation}
For each signal channel, the background  is composed of oscillating \Bd\ mesons
with probability given by the expression~(\ref{eq:qjt})  
and of combinatorial background, with probability given by the reconstructed 
proper time distribution. For a fraction $f_j^k$ of the $j$ component 
($j= s, \; d,$ and combinatorial background $cb$) in the total sample of 
type $k$, one obtains the probability density: 
\begin{equation*}
\mathrm{pdf}_k(t,\mu) = 
\sum_{j=s,d,cb} f_j^k \left  [ (1-\omega_j) q_j(t,\mu) + \omega_j q_j(t,-\mu) 
                      \right ] 
\end{equation*}
The index $k=1$ denotes the $\Bs \to \Dsm \pip$ 
channel and $k=2$ the $\Bs \to \Dsm \aonep$ channel.
The likelihood of the total sample is written as
\begin{equation}
{\cal L}(\Delta m_s, \Delta \Gamma_s) = 
    \prod_{k=1}^{N_{\mathrm{ch}}^{ }} \;
    \prod_{i=1}^{N_{\mathrm{ev}}^k} \mathrm{pdf}_k(t_i,\mu_i)
\label{fun:LL}
\end{equation}
where $N_{\mathrm{ev}}^k$ is the total number of events of  type $k$, and 
$N_{\mathrm{ch}} = 2$. Each $\mathrm{pdf}_k$ was properly normalized to unity.

\section{Extraction of \dmsSec\ measurement limits and accuracy}
\label{dmsLimAcc}

The maximum value of \dms\  measurable in ATLAS was estimated using a 
simplified Monte Carlo model. The input parameters of this model were:
for each signal channel $k$ 
the number of signal events, $N_\mathrm{sig}^k$, the number of background 
events from  \Bd\ decays, $N_{\Bd}^k$, and the number 
of events for the combinatorial background, $N_\mathrm{cb}^k$; 
the characteristics of the events involved in the computation of the 
proper-time resolution (see below); the wrong-tag fraction. 
The wrong-tag fraction was assumed to be the same 
for both \Bs\ and \Bd\ mesons:  $\omega_s = \omega_d =0.22$
(see Ref.~\cite{wTag}). 
A Monte Carlo sample with  $N_\mathrm{sig}=N_\mathrm{sig}^1+N_\mathrm{sig}^2$ 
signal events oscillating with a given frequency \dms\, together with 
$N_{\Bd}=N_{\Bd}^1 + N_{\Bd}^2$  
background events oscillating with frequency  \dmd\ and  
$N_\mathrm{cb}=N_\mathrm{cb}^1+N_\mathrm{cb}^2$ combinatorial events 
(no oscillations), was generated in the following way. 
For each event with an oscillating $b$ hadron, the true proper time  was 
generated according to an exponential distribution with the slope
$[- \Gamma (1 \pm \Delta \Gamma / (2 \Gamma))]^{-1}$ which takes into account
the contribution of a non-zero $\Delta \Gamma$. Here $\Gamma = \Gamma_s$,
$\Delta \Gamma = \Delta \Gamma_s$ for \Bs\ events and  $\Gamma = \Gamma_d$,
$\Delta \Gamma =0$ for \Bd\ events. The sign in the slope is chosen according 
to the contribution of each proper time, corresponding to   
$\Gamma (1 \pm \Delta \Gamma / (2 \Gamma))$, to the exponential decay from 
Eq.~(\ref{eq:oscPr}), 
$\mathrm{e}^{- \Gamma_{j} t_0} \cosh (\Delta\Gamma_{j} t_0 /2)$,
where the $\cosh$ term was factorized from the expression in 
parentheses and included in the exponential part.
The uncertainty on the measurement of the transverse decay length,
$\sigma_{d_{xy}}$, and the true value of the g-factor, $g_0$, were generated 
at random according to the distributions obtained from the simulated samples, 
fitted with appropriate combinations of Gaussian and exponential functions. 
From the computed true decay length, $d_{xy}^0=t_0/g_0$, the corresponding 
reconstructed decay length was generated as 
$d_{xy} = d_{xy}^0 + S_{d_{xy}} \sigma_{d_{xy}} \Omega$. The reconstructed 
g-factor was generated as $g = g_0 + g_0 S_g \Omega^\prime$. Both $\Omega$
and $\Omega^\prime$ are random numbers distributed according to the normal 
distribution. 
From the transverse decay length and g-factor, the reconstructed proper time 
was then computed as $t = g d_{xy}$. The probability for the event to be 
mixed or unmixed was determined from the $t_0$ and  \dms\ values (\dmd\ value
if the event was a \Bd\ event) using the expression 
$\left (1 - \cos(\Delta m_{j} t_0) /\cosh (\Delta\Gamma_{j} t_0 /2) \right)/2$ 
left out from
Eq.~(\ref{eq:oscPr}) after the exponential part is extracted. 
For a fraction of the events, selected at random, the state was changed 
between mixed and unmixed, according to the wrong-tag fraction, 
$\omega_\mathrm{tag}$. 
For the combinatorial background, the reconstructed proper time was generated 
assuming that it has the same distribution as the one for \Bs\ mesons. 
Half of the combinatorial events were added to the mixed events and half to 
the unmixed events. 

\subsection{\dmsSec\ measurement limits}
\label{dmsLim}

The \dms\ measurement limits were obtained applying the amplitude fit 
method~\cite{AFit} to the `data sample' generated as above. In this method
a new parameter, the \Bs\ oscillation amplitude ${\cal A}$, is introduced in 
the likelihood function by replacing the term `$\mu_0 \cos\Delta m_{s} t_0$' 
with `$\mu_0 {\cal A} \cos\Delta m_{s} t_0$' in the \Bs\ probability density 
function given by Eq.~(\ref{eq:oscPr}).  
For each value of \dms, the new likelihood function is minimized with respect
to ${\cal A}$, keeping all other parameters fixed, and a value 
${\cal A} \pm \sigma_{\cal A}^{\mathrm{stat}} $ is obtained. 
One expects, within the estimated uncertainty,  ${\cal A} = 1$ for \dms\ 
close to its true value, and ${\cal A} = 0$ for \dms\ far from the true value.
One defines a $5\sigma$ measurement limit as the value of \dms\ for which
$1/\sigma_{\cal A} =5$, and a sensitivity at 95\% confidence limit as the 
value of \dms\ for which $1/\sigma_{\cal A} = 1.645$. Limits are computed 
with the statistical uncertainty  $\sigma_{\cal A}^{\mathrm{stat}}$, and, 
in some cases, with the total uncertainty 
$\sigma_{\cal A}^{\mathrm{total}} = \sigma_{\cal A}^{\mathrm{stat}} + 
                                    \sigma_{\cal A}^{\mathrm{syst}}$.
The systematic uncertainty $\sigma_{\cal A}^{\mathrm{syst}}$ is described 
in the next section. 

\subsection{Systematic uncertainties}
\label{dmsSys}

The systematic uncertainties on the \Bs\ oscillation amplitude 
$\sigma_{\cal A}^{\mathrm{syst}}$ are computed as
\begin{equation}
  \label{eq:Asys}
  \sigma_{\cal A}^{\mathrm{syst}} = \Delta {\cal A} + 
         (1-{\cal A}) \D \frac{\Delta \sigma_{\cal A}^{\mathrm{stat}}}
                           {\sigma_{\cal A}^{\mathrm{stat}}}
\end{equation}
where $\Delta {\cal A}$ is the difference between the value of the amplitude
when a single parameter is changed and the analysis repeated, and the value 
for the `nominal' set of parameters; 
$\Delta \sigma_{\cal A}^{\mathrm{stat}}$ is defined in a similar way. 

The following contributions to the systematic uncertainties were considered:
\begin{itemize}
\item A relative error of 5\% was considered for the wrong-tag fraction 
  for both \Bs\ and \Bd. 
\item The widths of the Gaussian functions from  the parametrization 
  of the proper time resolution given in Table~\ref{tab:ptres2G_par} 
  were varied by $\pm 1\sigma$.
\item The fraction $f_{\Bs} = BR (\bar{b} \to \Bs)$, the \Bs\ lifetime and 
  the \dmd\ value were varied separately  by the 
  uncertainty quoted in Ref.~\cite{PDG1998}. 
\item An uncertainty of 5\%  was assumed for the decay time
  $\tau_{\mathrm{cb}}$ of the
  combinatorial background. The shape remained exponential,
  only the decay time was modified.
\end{itemize}

These contributions were added in quadrature to give the systematic 
uncertainty. 

Table~\ref{tab:syst_dms} shows the dependence of the 
amplitude and its statistical and systematic uncertainties on \dms, 
as well as the contribution of each component to the systematic 
uncertainty for an integrated luminosity of 10~\ifb. In the generated event
samples, the value of $\dms$ was set to $\dms^{\mathrm{gen}} = \infty$.

\begin{table*}[htbp]
  \begin{center}
    \begin{tabular}{lrrrrrrrrr} 
\hline
\multicolumn{1}{c}{\dms}      &  
   0 \ips\ &  
   5 \ips\ & 
  10 \ips\ & 
  15 \ips\ &  
  20 \ips\ & 
  25 \ips\ &
  30 \ips\ &  
  35 \ips\ & 
  40 \ips\ \\
\hline
\multicolumn{1}{c}{${\cal A}$} &  
$ 0.0450$  &
$-0.1146$  &          
$ 0.1892$  &          
$-0.0337$  &          
$ 0.0420$  &          
$ 0.0072$  &
$-0.2906$  &
$ 0.4318$  & 
$-0.5406$  \\[0.75ex]
\multicolumn{1}{c}{$\sigma_{\cal A}^{\mathrm{stat}}$} &  
$\pm 0.0476$  &         
$\pm 0.0707$  &
$\pm 0.0895$  &
$\pm 0.1183$  &
$\pm 0.1673$  &
$\pm 0.2456$  &
$\pm 0.3570$  &
$\pm 0.5640$  & 
$\pm 0.9300$  \\[0.75ex]
\multicolumn{1}{c}{$\sigma_{\cal A}^{\mathrm{syst}}$} &  
{\large $^{+0.0967}_{-0.0838}$} &
{\large $^{+0.1285}_{-0.1018}$} &
{\large $^{+0.1297}_{-0.0960}$} &
{\large $^{+0.1468}_{-0.1144}$} &
{\large $^{+0.1796}_{-0.1424}$} &
{\large $^{+0.2035}_{-0.1580}$} &
{\large $^{+0.2980}_{-0.2264}$} &
{\large $^{+0.3674}_{-0.2658}$} &
{\large $^{+0.3636}_{-0.2592}$} \\
\hline
Systematic contributions & & & & & & & & &   \\[1.75ex]
 - wrong-tag fraction &
{\large $^{+0.0387}_{-0.0382}$} &         
{\large $^{+0.0361}_{-0.0360}$} &          
{\large $^{+0.0445}_{-0.0349}$} &          
{\large $^{+0.0389}_{-0.0391}$} &          
{\large $^{+0.0389}_{-0.0312}$} &          
{\large $^{+0.0394}_{-0.0329}$} &
{\large $^{+0.0408}_{-0.0342}$} &
{\large $^{+0.0355}_{-0.0319}$} & 
{\large $^{+0.0418}_{-0.0364}$} \\[1.75ex] 
 - proper time resolution &
{\large $^{+0.0000}_{-0.0000}$} &
{\large $^{+0.0126}_{-0.0127}$} &
{\large $^{+0.0551}_{-0.0450}$} &
{\large $^{+0.0950}_{-0.0758}$} &
{\large $^{+0.1359}_{-0.1143}$} &
{\large $^{+0.1700}_{-0.1362}$} &
{\large $^{+0.2751}_{-0.2098}$} &
{\large $^{+0.3469}_{-0.2490}$} &
{\large $^{+0.3458}_{-0.2425}$} \\[1.75ex] 
 - $f_s$ fraction  &
{\large $^{+0.0884}_{-0.0742}$} &
{\large $^{+0.1227}_{-0.0943}$} &
{\large $^{+0.1086}_{-0.0773}$} &
{\large $^{+0.1049}_{-0.0761}$} &
{\large $^{+0.1107}_{-0.0790}$} &
{\large $^{+0.1035}_{-0.0724}$} &
{\large $^{+0.1069}_{-0.0780}$} &
{\large $^{+0.1146}_{-0.0861}$} &
{\large $^{+0.0888}_{-0.0661}$} \\[1.75ex] 
 - \Bs\ lifetime   &
{\large $^{+0.0038}_{-0.0036}$} &
{\large $^{+0.0009}_{-0.0009}$} &
{\large $^{+0.0034}_{-0.0032}$} &
{\large $^{+0.0030}_{-0.0033}$} &
{\large $^{+0.0060}_{-0.0000}$} &
{\large $^{+0.0146}_{-0.0076}$} &
{\large $^{+0.0010}_{-0.0014}$} &
{\large $^{+0.0132}_{-0.0088}$} & 
{\large $^{+0.0372}_{-0.0390}$} \\[1.75ex] 
 - \dmd\  &
{\large $^{+0.0055}_{-0.0055}$} &
{\large $^{+0.0001}_{-0.0001}$} &
{\large $^{+0.0003}_{-0.0003}$} &
{\large $^{+0.0004}_{-0.0003}$} &
{\large $^{+0.0002}_{-0.0002}$} &
{\large $^{+0.0007}_{-0.0007}$} &
{\large $^{+0.0001}_{-0.0002}$} &
{\large $^{+0.0003}_{-0.0003}$} &
{\large $^{+0.0020}_{-0.0019}$} \\[1.75ex] 
 - $\tau_{\mathrm{cb}}$  &
{\large $^{+0.0012}_{-0.0013}$} &
{\large $^{+0.0018}_{-0.0016}$} &
{\large $^{+0.0014}_{-0.0021}$} &
{\large $^{+0.0031}_{-0.0027}$} &
{\large $^{+0.0001}_{-0.0002}$} &
{\large $^{+0.0057}_{-0.0061}$} &
{\large $^{+0.0047}_{-0.0041}$} &
{\large $^{+0.0094}_{-0.0109}$} &
{\large $^{+0.0403}_{-0.0339}$} \\[1.75ex] 
\hline
    \end{tabular}
    \caption{The oscillation amplitude ${\cal A}$ and its statistical and 
      systematic uncertainties as a function of \dms\ for an integrated 
      luminosity of 10~\ifb. The contribution of each 
      component to the systematic uncertainty is also given.}    
    \label{tab:syst_dms}
  \end{center}
\end{table*}

From Table~\ref{tab:syst_dms}, it can be seen that the dominant
contributions to the systematic uncertainty come from the uncertainty
on  $f_{\Bs}$ fraction and from  the parametrization of the proper
time resolution. 

The uncertainty on  $f_{\Bs}$ fraction has the value
given in Ref.~\cite{PDG1998}. It is, however, expected that the
uncertainty will be much smaller at the time this analysis will be
done with data. Even now (October 2001), the most up-to-date preliminary 
value~\cite{BoscWG} of $f_{\Bs}$ is $0.099 \pm 0.011$, the uncertainty
being at the level of 11\%, to be compared with the value of 17\% used
here for the consistency of the data. If one considers the intense 
activity in $B-$physics, one  can assume that an uncertainty of
$\sim 5\%$ will be achieved at the time of data analysis.

The uncertainties on the parameters from  the parametrization  of the
proper time resolution depend on the MC statistics. A larger MC sample can
be generated than the one used in this work, in order to reduce these
uncertainties. It is therefore reasonable to assume that the uncertainty
on the widths from Table ~\ref{tab:ptres2G_par} can be reduced to half
of the actual values.

The systematic uncertainties computed with these `projected'
uncertainties are given in Table~\ref{tab:syst_dms_pro}. For the other
contributions, the values from Table~\ref{tab:syst_dms} were used. 

\begin{table*}[htbp]
  \begin{center}
    \begin{tabular}{lrrrrrrrrr} 
\hline
\multicolumn{1}{c}{\dms}      &  
   0 \ips\ &  
   5 \ips\ & 
  10 \ips\ & 
  15 \ips\ &  
  20 \ips\ & 
  25 \ips\ &
  30 \ips\ &  
  35 \ips\ & 
  40 \ips\ \\
\hline
\multicolumn{1}{c}{${\cal A}$} &  
$ 0.0450$  &
$-0.1146$  &          
$ 0.1892$  &          
$-0.0337$  &          
$ 0.0420$  &          
$ 0.0072$  &
$-0.2906$  &
$ 0.4318$  & 
$-0.5406$  \\[0.75ex]
\multicolumn{1}{c}{$\sigma_{\cal A}^{\mathrm{stat}}$} &  
$\pm 0.0476$  &         
$\pm 0.0707$  &
$\pm 0.0895$  &
$\pm 0.1183$  &
$\pm 0.1673$  &
$\pm 0.2456$  &
$\pm 0.3570$  &
$\pm 0.5640$  & 
$\pm 0.9300$  \\[0.75ex]
\multicolumn{1}{c}{$\sigma_{\cal A}^{\mathrm{syst}}$} &  
{\large $^{+ 0.0494}_{-0.0492}$}&
{\large $^{+ 0.0564}_{-0.0471}$}&
{\large $^{+ 0.0596}_{-0.0475}$}&
{\large $^{+ 0.0679}_{-0.0617}$}&
{\large $^{+ 0.0848}_{-0.0661}$}&
{\large $^{+ 0.0976}_{-0.0823}$}&
{\large $^{+ 0.1370}_{-0.1167}$}&
{\large $^{+ 0.1660}_{-0.1411}$}&
{\large $^{+ 0.1747}_{-0.1496}$}\\
\hline
Systematic contributions & & & & & & & & &   \\[1.75ex]
 - proper time resolution &
{\large $^{+ 0.0000}_{-0.0000}$}&
{\large $^{+ 0.0134}_{-0.0133}$}&
{\large $^{+ 0.0229}_{-0.0226}$}&
{\large $^{+ 0.0476}_{-0.0379}$}&
{\large $^{+ 0.0681}_{-0.0532}$}&
{\large $^{+ 0.0825}_{-0.0708}$}&
{\large $^{+ 0.1277}_{-0.1077}$}&
{\large $^{+ 0.1579}_{-0.1341}$}&
{\large $^{+ 0.1586}_{-0.1340}$}\\[1.75ex] 
 - $f_s$ fraction  &
{\large $^{+ 0.0299}_{-0.0301}$}&
{\large $^{+ 0.0412}_{-0.0272}$}&
{\large $^{+ 0.0321}_{-0.0226}$}&
{\large $^{+ 0.0285}_{-0.0287}$}&
{\large $^{+ 0.0316}_{-0.0238}$}&
{\large $^{+ 0.0305}_{-0.0242}$}&
{\large $^{+ 0.0278}_{-0.0286}$}&
{\large $^{+ 0.0333}_{-0.0265}$}&
{\large $^{+ 0.0252}_{-0.0205}$}\\[1.75ex] 
\hline
    \end{tabular}
    \caption{The oscillation amplitude ${\cal A}$ and its statistical and 
      systematic uncertainties as a function of \dms\ for an integrated 
      luminosity of 10~\ifb, computed with the `projected
      uncertainties' on proper time parametrization and on $f_s$.
      See the text for details.}    
    \label{tab:syst_dms_pro}
  \end{center}
\end{table*}

The value of the systematic uncertainty as computed here should be considered 
with caution, as a rough estimate. 

\subsection{Results}
\label{results}

The amplitude as a function of \dms\ for the nominal set of parameters
defined in the previous sections, $\dgs = 0$ and an integrated  luminosity 
of 10~\ifb\ is shown in Fig.~\ref{fig:ampl_dms}. 
Fig.~\ref{fig:sign_dms} shows the significance of the measurement in units 
of $\sigma_{\cal A}$.
The $5\sigma$ measurement limit is 22.5~\ips\ and the 95\% CL sensitivity
is 36.0~\ips, when computed with the statistical uncertainty only.
Computed with the total uncertainty, the $5\sigma$ measurement limit is 
16.0~\ips\ and the 95\% CL sensitivity is 34.5~\ips\ for the actual
systematic uncertainties, and 21~\ips\ and 35.5~\ips\ for the
projected systematic uncertainties. 

For an integrated luminosity of 30~\ifb, the $5\sigma$ measurement 
limit is 29.5~\ips\ and the 95\% CL sensitivity is 41.0~\ips, computed
with the statistical uncertainty only. The $5\sigma$ measurement
limit and the 95\% CL sensitivity become 18.5~\ips\ and 37.5~\ips, 
respectively, for the actual set of systematic uncertainties and
27.0~\ips\ and 40.5~\ips\ for the projected systematic uncertainties.

\begin{figure}[htbp] 
  \centering
  \includegraphics[width=90mm,height=75mm]{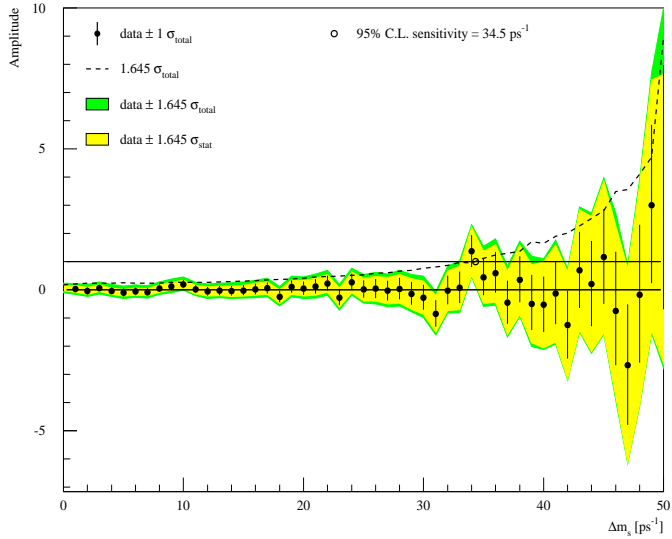}
  \caption{The \Bs\ oscillation amplitude as a function of \dms\ for 
           an integrated luminosity of 10~\ifb.} 
  \label{fig:ampl_dms}  
\end{figure}

\begin{figure} 
  \centering
  \includegraphics[width=90mm,height=75mm]{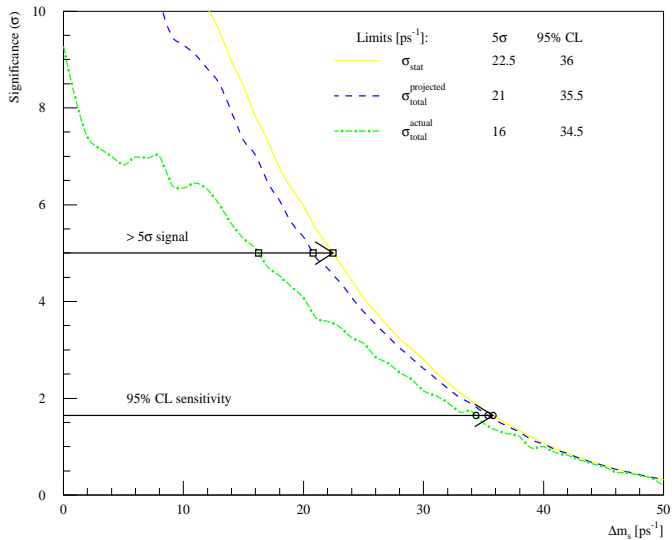}
  \caption{The measurement significance as a function of \dms\ for 
           an integrated luminosity of 10~\ifb.} 
  \label{fig:sign_dms}  
\end{figure}

\section{Dependence of \dmsSec\ measurement limits on experimental quantities}
\label{dmsDependence}

Some quantities involved in the measurement of the \Bs\ oscillations 
are not yet known with enough precision.  The cross section for $b \bar{b}$
production  could be more than twice the assumed cross section~\cite{PhysTDR};
some decay branching ratios are also not well determined. 
The value of \dgs\ is not yet measured. 
The characteristics of the detector could be slightly different, 
depending on the final configuration of the detector.
The shape and the fraction of the background will depend, among other factors,
on the accelerator luminosity and on the characteristics of the detector.
It is therefore necessary to estimate the dependence of the measurement 
sensitivity %and of the expected accuracy 
for various values of the parameters.
The limits presented in this section are computed with the statistical 
error only.

The dependence of the \dms\ measurement limits on \dgsg\ was determined 
for an integrated luminosity of 30~\ifb, other parameters having the  
nominal value. The \dgsg\ was used as a fixed parameter in the amplitude 
fit method. The results are shown in Fig.~\ref{fig:dms_dg} and the numerical
values are  given in Table~\ref{tab:dms_dg}. No sizeable effect is seen up 
to $\dgsg < 30\%$, therefore \dgs\ was set to zero for all other cases. 

\begin{table}[htbp]
  \begin{center}
    \begin{tabular}{ccc} 
\hline
\multicolumn{1}{c}{\dgsg} & $5 \sigma$ limit & $95\%$ CL sensitivity  \\ 
\multicolumn{1}{c}{(\%)}  &   (\ips)               &  (\ips)          \\ 
\hline
\phantom{10}0             &   29.5                 &   41.0           \\ 
\phantom{1}10             &   29.5                 &   41.0           \\
\phantom{1}20             &   29.5                 &   41.0           \\
\phantom{1}30             &   29.0                 &   41.0           \\ 
\phantom{1}40             &   29.0                 &   40.5           \\ 
\phantom{1}80             &   27.0                 &   39.0           \\ 
          100             &   25.5                 &   38.0           \\ 
\hline
    \end{tabular}
    \caption{The dependence of \dms\ measurement limits on \dgsg.}    
    \label{tab:dms_dg}
  \end{center}
\end{table}

\begin{figure} 
  \centering
  \includegraphics{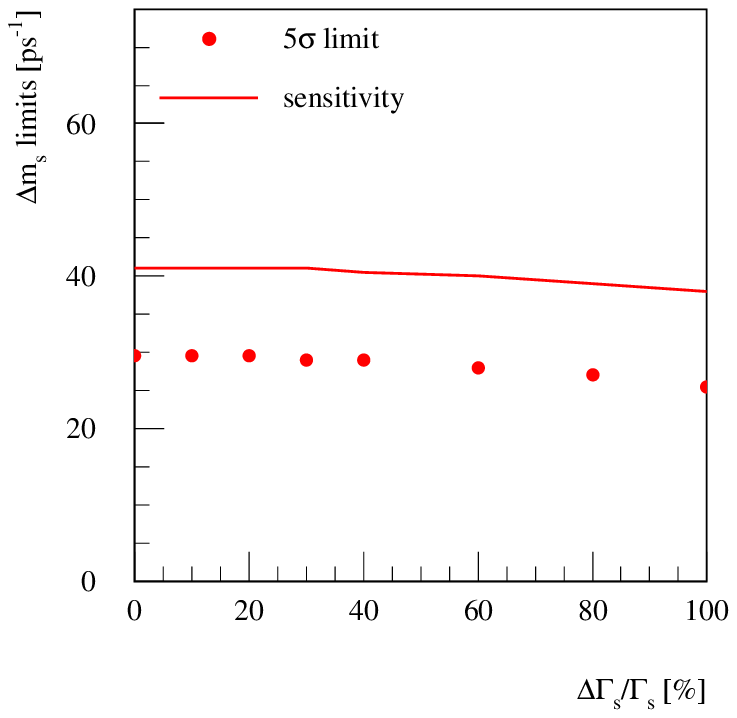}
  \caption{The dependence of \dms\ measurement limits on \dgsg.} 
  \label{fig:dms_dg}  
\end{figure}

The dependence of the \dms\ measurement limits on the integrated luminosity
is shown in Fig.~\ref{fig:dms_lum}, with the numerical
values given in Table~\ref{tab:dms_lum}. 

\begin{figure}
  \centering
  \includegraphics{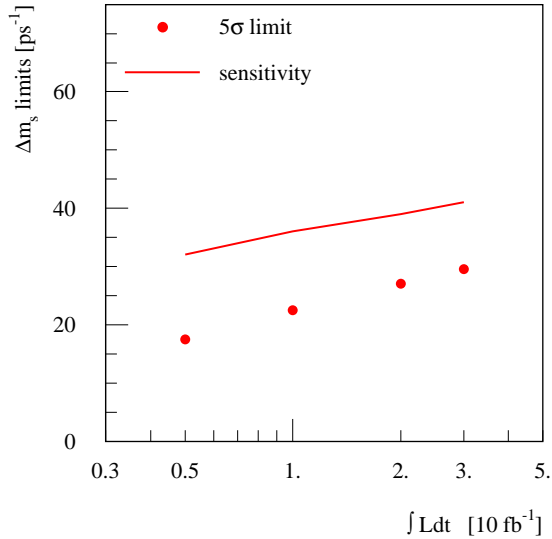}
  \caption{The dependence of \dms\ measurement limits on the 
           integrated luminosity.}    
  \label{fig:dms_lum}  
\end{figure}

\begin{table}[htbp]
  \begin{center}
    \begin{tabular}{ccc} 
\hline
Luminosity   & $5 \sigma$ limit         &   $95\%$ CL sensitivity  \\ 
  (\ifb)     &   (\ips)                 &          (\ips)          \\ 
\hline
\phantom{1}5 &     17.5                 &      32.0                \\          
          10 &     22.5                 &      36.0                \\       

          20 &     27.0                 &      39.0                \\  
          30 &     29.5                 &      41.0                \\       
\hline
    \end{tabular}
    \caption{The dependence of \dms\ measurement limits on the 
             integrated luminosity.}    
    \label{tab:dms_lum}
  \end{center}
\end{table}

The shape and the fraction of the combinatorial background were varied 
within reasonable values. It was assumed that the shape remains exponential
and only the decay time was modified. The number of events for the 
combinatorial background was varied within $\pm 50\%$ of the number 
determined in Section~\ref{EvSel}, keeping the number of \Bs\ and \Bd\ 
events to the nominal values. The dependence of the \dms\ measurement 
limits on the decay time of the combinatorial background
is shown in Fig.~\ref{fig:dms_cbg}, with the numerical
values given in Table~\ref{tab:dms_cbg}, while the dependence on the number
of events is given in Fig.~\ref{fig:dms_Ncbg} and Table~\ref{tab:dms_Ncbg}.

\begin{figure}[htbp] 
  \centering
  \includegraphics{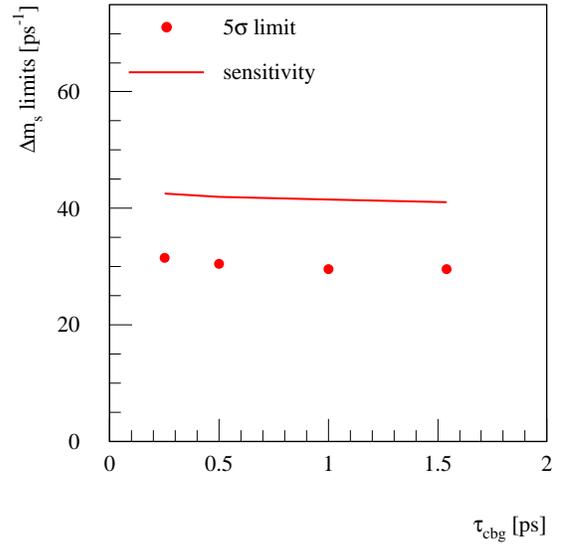}
  \caption{The dependence of \dms\ measurement limits on the 
           decay time of the combinatorial background, assuming  
           an exponential decay.}    
  \label{fig:dms_cbg}  
\end{figure}

\begin{table}[htbp]
  \begin{center}
    \begin{tabular}{ccc} 
\hline
$\tau_\mathrm{cbg}$ & $5 \sigma$  limit   &   $95\%$ CL sensitivity  \\ 
  (ps)     &   (\ips)                 &          (\ips)              \\
\hline
0.25                &   31.5             &      42.5                 \\       
0.50                &   30.5             &      42.0                 \\ 
1.00                &   29.5             &      41.5                 \\ 
1.54                &   29.5             &      41.0                 \\ 
\hline
    \end{tabular}
    \caption{The dependence of \dms\ measurement limits on the 
             decay time of the combinatorial background, assuming  
             an exponential decay.}    
    \label{tab:dms_cbg}
  \end{center}
\end{table}

\begin{figure}[htbp] 
  \centering
  \includegraphics{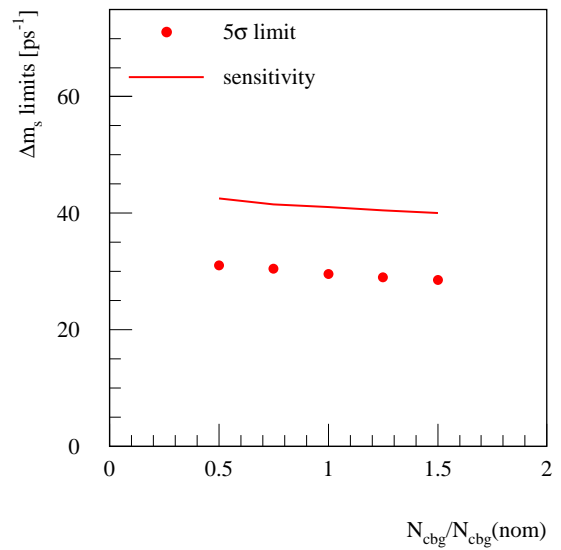}
  \caption{The dependence of \dms\ measurement limits on the 
           number of events for the combinatorial background.}    
  \label{fig:dms_Ncbg}  
\end{figure}

\section{Accuracy of the \dmsSec\ measurement}
\label{dmsAcc}

\begin{table}[htbp]
  \begin{center}
    \begin{tabular}{ccc} 
\hline
$N_\mathrm{cbg}/N_\mathrm{cbg}^\mathrm{nom}$ & 
$5 \sigma$  limit                            &   
$95\%$ CL sensitivity                        \\ 
                    &   (\ips)       &          (\ips)     \\
\hline
0.50                &   31.0         &      42.5           \\       
0.75                &   30.5         &      41.5           \\        
1.00                &   29.5         &      41.0           \\       
1.25                &   29.0         &      40.5           \\       
1.50                &   28.5         &      40.0           \\       
\hline
    \end{tabular}
    \caption{The dependence of \dms\ measurement limits on the 
             number of events for the combinatorial background.}    
    \label{tab:dms_Ncbg}
  \end{center}
\end{table}

For \dms\ values smaller than the $5\sigma$ measurement limit, the expected 
accuracy is estimated using the  log-likelihood method, with the likelihood 
function given by Eq.~(\ref{fun:LL}). In the fit, the \dms\ value was free, 
while the other parameters were fixed to their nominal values. An example 
of the likelihood function is given in Figure~\ref{fig:LL_10_225}  
for an integrated luminosity of 10~\ifb\ and a generated 
$\dms^{\mathrm{gen}} = 22.5$~\ips.

\begin{figure} 
  \centering
  \includegraphics[width=90mm,height=75mm]{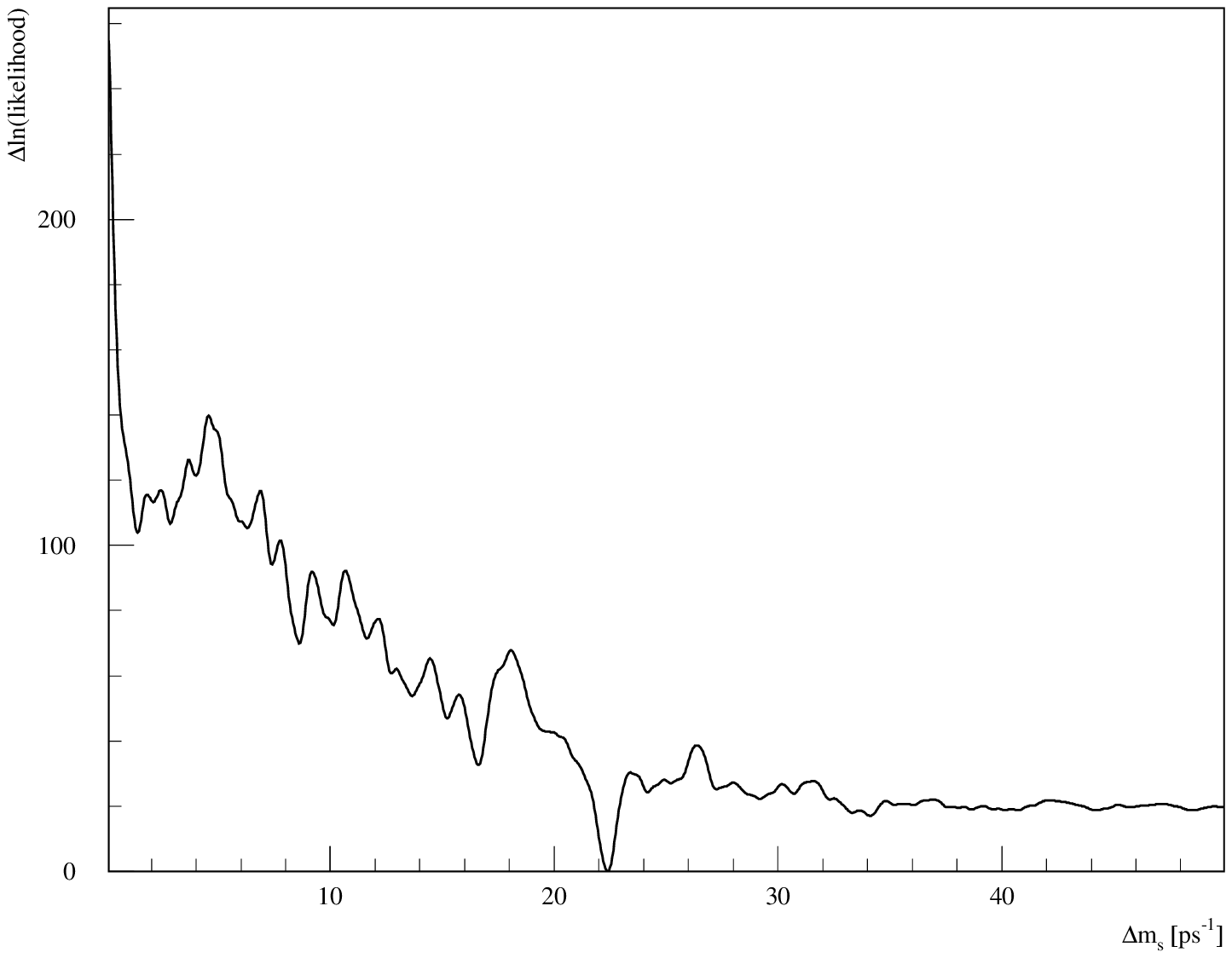}
  \caption{The negative log-likelihood function for an integrated          
           luminosity of 10~\ifb\ and a generated 
           $\dms^{\mathrm{gen}} = 22.5$~\ips.}    
  \label{fig:LL_10_225}  
\end{figure}

The accuracy was determined for different values of the integrated luminosity.
The results are given in Table~\ref{tab:dmsAcc}.

The systematic accuracy of the \dms\ measurement was obtained similarly to
the systematic uncertainty on the measurement limits: one parameter was
changed, keeping the other fixed to their nominal values, and the analysis 
was repeated. The difference between the value of the fitted \dms\ parameter
when a single parameter is changed and the analysis repeated, and the value 
for the `nominal' set of parameters was taken as the systematic contribution 
corresponding to that parameter. The total systematic uncertainty is the 
quadratic sum of the individual contributions. The parameters changed are 
the same as for the measurement limits, see Section~\ref{dmsSys}, within 
the `actual' uncertainties.

\begin{table}[htbp]
  \begin{center}
    \begin{tabular}{cccc} 
\hline
Luminosity                                      &
$\dms^{\mathrm{gen}}$                           & 
$\dms^{\mathrm{rec}} \pm \sigma^{\Delta m_s}_{\mathrm{stat}} 
                     \pm \sigma^{\Delta m_s}_{\mathrm{syst}}$  &   
Obs.                                                        \\ 
  (\ifb)        & (\ips) &      (\ips)                  &   \\ 
\hline
\phantom{1} 5   & $17.5$ & $17.689 \pm 0.083 \pm 0.002$   & $5 \sigma$ limit \\
\hline 
                & $15.0$ & $15.021 \pm 0.049 \pm 0.002$   & \\ %\cline{2-3}
\raisebox{1.5ex}[0cm][0cm]{10}   
                & $22.5$ & $22.396 \pm 0.072 \pm 0.005$   & $5 \sigma$ limit \\
\hline
                & $15.0$ & $14.949 \pm 0.033 \pm 0.002$   & \\ %\cline{2-3}
           20   & $20.0$ & $20.041 \pm 0.068 \pm 0.005$   & \\ %\cline{2-3}
                & $27.0$ & $26.948 \pm 0.070 \pm 0.003$   & $5 \sigma$ limit \\
\hline
                & $15.0$ & $14.942 \pm 0.028 \pm 0.004$   & \\ %\cline{2-3}
           30   & $20.0$ & $20.010 \pm 0.043 \pm 0.002$   & \\ %\cline{2-3}
                & $29.5$ & $29.708 \pm 0.083 \pm 0.007$   & $5 \sigma$ limit \\
\hline
    \end{tabular}
    \caption{The accuracy of \dms\ measurement as a function of the 
             integrated luminosity. $\sigma^{\Delta m_s}_{\mathrm{stat}}$
             represents the statistical uncertainty, 
             $\sigma^{\Delta m_s}_{\mathrm{stat}}$ the systematic uncertainty.}
    \label{tab:dmsAcc}
  \end{center}
\end{table}

One can see from Table~\ref{tab:dmsAcc} that the \dms\ measurement, if within 
the reach of the ATLAS detector, will be dominated by the statistical errors.

\section{Conclusions}
\label{conclusion}

In this note, the performance of the ATLAS detector to measure the \Bs\ 
oscillations was estimated using Monte Carlo simulated events, propagated 
through a detailed simulation of the detector. For an integrated luminosity 
of 10~\ifb, a $5\sigma$ measurement will be possible up to 22.5~\ips, 
the experiment being 95\% CL sensitive up to 36.0~\ips.
These limits increase to 29.5~\ips\ and 41.0~\ips, respectively, for an 
integrated luminosity of 30~\ifb. The effect of the systematic uncertainties 
on the measurement limits is rather limited, if the precision of the
$f_{\Bs}$ fraction is improved with respect to the actual value,
and if enough Monte Carlo events are generated for the parametrization of the 
proper time.

If the \dms\ value will be within the ATLAS reach, the measurement will be 
dominated by the statistical error. A total uncertainty of the order of 
$\sim 0.07$~\ips\ is expected for a value of $\dms = 22.5$~\ips\ for an 
integrated luminosity of 10~\ifb, the total uncertainty decreasing to 
$\sim 0.04$~\ips\ for a luminosity of 30~\ifb. 

The dependence of the measurement sensitivity on various parameters which 
will be known only at the time of the data analysis was also evaluated.

The values obtained in this note for the measurement limits and accuracies 
should be re-evaluated at a later time, taking into account the changes in 
the detector geometry and in the simulation and reconstruction software.

\begin{acknowledgement}
{\it Acknowledgements.}
This work has been performed within the ATLAS collaboration, and we thank
collaboration members for helpful discussions. We have made use of the 
physics analysis framework and tools which are the result of 
collaboration-wide effort. The authors would like to thank P.~Eerola and
N.~Ellis for fruitful discussions, and  E.~Kneringer for computing 
support. The work was supported by the Federal Ministry of Education, Science 
and Culture, Austria.   
\end{acknowledgement}

\end{document}